%Paper: patt-sol/9404006
%From: Yuji Kodama <kodama@math.ohio-state.edu>
%Date: Wed, 27 Apr 1994 16:54:46 -0400

\input amstex
\documentstyle{amsppt}
\magnification=1200
\NoBlackBoxes
\def\diag{\text{diag}}

\def\ker{\text{ker}}

\def\list#1#2{\noindent\hangindent 3em\hangafter1\hbox to 3em{\hfil#1\quad}#2}

\def\rbx{\hfill{\vbox{\hrule\hbox{\vrule\kern6pt\vbox{\kern6pt}\vrule}\hrule}}}
\def\bx{\vbox{\hrule\hbox{\vrule\kern6pt\vbox{\kern6pt}\vrule}\hrule}}
\def\lmatrix#1{\null\,\vcenter{\normalbaselines\mathsurround0pt\ialign
{$##$\hfill&&\quad$##$\hfil\crcr\mathstrut\crcr\noalign{\kern-\baselineskip}
#1\crcr\mathstrut\crcr\noalign{\kern-\baselineskip}}}\,}
\rightheadtext{NORMAL FORM, SYMMETRY AND INFINITE DIMENSIONAL LIE ALGEBRA}

\topmatter
\title Normal Form, Symmetry and Infinite Dimensional\\
Lie Algebra for System of ODE's
\endtitle
\author Yuji Kodama
\endauthor
\address Department of Mathematics, The Ohio State University, Columbus, OH
43210
\endaddress
\abstract{The normal form for a system of ode's is constructed from its
 polynomial
symmetries of the linear part of the system, which is assumed to be
semi-simple. The symmetries are shown to have a simple structure such as
 invariant function times symmetries of degree one called basic
symmetries.  We also show that the set of symmetries naturally forms
an infinite dimensional Lie algebra graded by the degree of invariant
 polynomials. This implies that if this algebra is non-commutative
then the method
of multiple scales with more than two scaling variables fails to apply.}
\endabstract
\endtopmatter

\document

\heading{1.}
\endheading

The method of normal form is one of the most important tools in the study of
nonlinear differential
equations [1].  It classifies the vector field near the critical point, and the
normal form consists
of the resonant vectors characterized by the linear part of the vector field.

Here we study the normal form on $\Bbb C^N$ where the linear part of the vector
field is assumed to
be semi-simple (diagonalizable) for a simplicity (the general case will be
discussed elsewhere).  The main purpose of this letter is to give a method to
determine
the normal form in terms of the symmetries of the linear part of the system.
We also show that for
certain cases the set of the symmetries forms an infinite dimensional Lie
algebra graded by the degree
of invariant polynomials.
As a consequence of this algebra structure, we see that the method of
multiple scales with more than two variables fails in general.
\heading{2.}
\endheading

Let us consider a system of equations for $x\in\Bbb C^N$,
$$\eqalignno{{dx \over dt} &= X_F(x) = F(x). &(1)\cr}$$
where the vector field $X_F$ is given by
$$\eqalignno{X_F &=\sum^N_{i=1}f_i(x){\partial \over \partial x_i} :=
F(x)\cdot{d \over dx}\,.
&(2)\cr}$$
The components of $X_F$, denoted by the lower-case letters $\{
f_i(x)\}^N_{i=1}$, are assumed to be
$$\eqalignno{f_i(x) &\in C^\infty (\Omega) &(3)\cr}$$
for some neighborhood $\Omega$ of a critical point $x_0$.  We assume $x_0=0$,
i.e., $X_0(0)=0$.  In order to classify the critical point, we write the system
in the series near the critical point,
$$\eqalignno{{dx \over dt} &= Ax+F^{(2)}(x)+\cdots, &(4)\cr}$$
where $F^{(\ell )}(x)$ is a vector valued function of homogeneous polynomial of
degree $\ell$.  Here we also assume that the matrix $A \in M_N(\Bbb C)$
 of the linear part is diagonalizable with the eigenvalues
$\lambda_1, \lambda_2, \dots, \lambda_N$. This case is refered to be
semi-simple. Now we consider
the normal form of (4) near $x=0$.  In his thesis, Poinc\'are showed:

\proclaim{Theorem} (Poinc\'are [1])  If there is no eigenvalue $\lambda_s$
satisfying
$$\eqalignno{\lambda_s &= \sum^N_{i=1}n_i\lambda_i &(5)\cr}$$
for some $n_i\in\Bbb Z_{\ge 0}$ and $|n|:=\Sigma n_i\ge 2$ (non-resonant
condition), then there exists a
formal change of variables $x=\phi (y)$ with $\phi (y)$ being polynomial, such
that the transformed system for $y$ is just the
linear part of (4),
$$\eqalignno{{dy \over dt} &= Ay\,. &(6)\cr}$$
\endproclaim

Thus the nonlinear parts in (4) can be eliminated by the non-resonant
condition, and the solution of
(4) near critical point can be described by that of the linear equation.  The
equation for $y$ is
called the {\it normal form} of (4).  For the case including the resonances, we
have:

\proclaim{Theorem} (Poinc\'are-Dulac [1])  The normal form of (4) is given by
$$\eqalignno{{dy \over dt} &= Ay+G^{(2)}(y)+\cdots, &(7)\cr}$$
where $G^{(\ell )}(y)$ is a vector valued function of homogeneous
polynomial of degree $\ell$, and is a solution of
$$\eqalignno{ L_A G^{(\ell )}(y) &= { \left\{I_N (Ay)\cdot {d \over dy} -
A\right\} } G^{(\ell)}(y)=0, &(8)\cr}$$
where $I_N$ is the $N\times N$ identity matrix.
\endproclaim

Outlines of the proofs of these theorems are as follows:

Under the change of coordinates, $x=\phi (y)$, the vector field $X_F$
becomes $X_G$ such that
$$ X_{\phi} \circ X_F = X_G \circ X_{\phi}. \eqno (9)$$
Then expanding $F, G$ and $\phi$ as
$$\eqalignno{ F(y) &= A y + F^{(2)}(y) + \cdots, \cr
  G(y) &= A y + G^{(2)}(y) + \cdots, \cr
 \phi (y) &= y + \phi^{(2)}(y) + \cdots, &(10)\cr}$$
(9) gives,
on the $n^{\text{th}}$ degree polynomial,
$$\eqalignno{L_A\phi^{(n)}&=
\widetilde{F}^{(n)}(y)-G^{(n)}(y),\quad &(11)\cr}$$
where $\widetilde{F}^{(n)}(y)$ is determined successively, e.g.,
$$\eqalign{\widetilde{F}^{(2)}(y) &= F^{(2)}(y)\,,\cr
\widetilde{F}^{(3)}(y) &= F^{(3)}(y)+\phi^{(2)}(y)\cdot{d \over
dy}F^{(2)}(y)-G^{(2)}(y)\cdot
{d \over dy}\phi^{(2)}(y)\,,\cr}$$
and so on.  In the case of $A=\diag (\lambda_1,\dots ,\lambda_N)$, the operator
$L_A$ in (8) is
$$\eqalignno{L_A &= I_N\left(\sum^N_{i=1}\lambda_iy_i{\partial \over \partial
y_i}\right) -A.  &(12)\cr}$$
 Then we first note that the eigenvectors of $L_A$
are given by a monomial of degree $|n|=\Sigma n_i$,
$$\eqalignno{\psi^{(n)}_\ell (y) &=
e_\ell\prod^N_{i=1}y^{n_i}_i~~~\quad~~~\text{for}~~\ell =1,\dots ,N
&(13)\cr}$$
with the eigenvalue $\Lambda^n_\ell$, $L_A\psi^{(n)}_\ell
=\Lambda^n_\ell\psi^{(n)}_\ell$,
$$\eqalignno{\Lambda^n_\ell &= \left(\sum^N_{i=1}\lambda_in_i\right)
-\lambda_\ell\,, &(14)\cr}$$
where $n_i\in\Bbb Z_{\ge 0}$ and $e_\ell $ is an eigenvector of $A$ with
eigenvalue $\lambda_\ell$,
$Ae_\ell =\lambda_\ell e_\ell$.  If the $\lambda_\ell$ satisfies the resonant
condition (5), that is
$\Lambda^n_\ell =0$, then the vector (13) is called the {\it resonant vector}.
Therefore, if there is
no resonance $(\ker L_A=\{0\})$, then one chooses $G^{(n)}(y)=0$, while for the
case of resonances, we
choose $G^{(n)}(y)\in\ker L_A\cap\widetilde{F}^{(n)}(y)$, which is the resonant
vector (13) of degree
$n$.
In the next section, we relate the resonant vector (13) to a symmetry of the
linear part of the
system (4).

\heading{3.}
\endheading

Let us first define the symmetry of the system (1).

\definition{Definition}  A vector field $X_S=S(x)\cdot{d \over dx}$ is a {\it
 symmetry} of (1), if it
commutes with the vector field $X_F$, i.e.
$$\eqalignno{[X_S,X_F]:&= X_S\circ X_F - X_F \circ X_S \cr
&=\left\{S(x)\cdot{dF(x) \over dx} - F(x)\cdot{dS(x) \over dx}\right\}\cdot{d
\over dx} = 0\,. &(15)\cr}$$
\enddefinition

The comdition (15) implies that the vector function $S(x)$ is a solution of the
linearized equation
of (1).  The symmetry can be found directly from the general solution of (1),
when the system is
integrable.  Namely we have:

\proclaim{Proposition 1}  Let $x(t)$ be the general solution of (1),
$$\eqalignno{x(t) &= U(t;C_1,\dots ,C_N) &(16)\cr}$$
with $N$ arbitrary parameters $C_i$.  Then the symmetries are obtained by
$$\eqalignno{S_i(x) &= {\partial U \over \partial C_i}(t;C_1,\dots ,C_N)\,.
&(17)\cr}$$
\endproclaim

\demo{Proof}
$$\eqalignno{{dS_i \over dt} &= {dx \over dt}\cdot {dS_i \over dx} = F(x)\cdot
{dS_i \over dx}\cr
&={\partial \over \partial C_i}\,{dx \over dt} = {\partial \over \partial C_i}
F= {\partial U \over
\partial C_i}\cdot{dF\over dx} = S_i(x)\cdot {dF\over dx}\,. &\bx\cr}$$
Thus the symmetry is a generator of the transformation group with $N$
parameters of the solution.
In (16), some of the parameters may be determined by the level of the invariant
functions of (1).
For example, an integrable Hamiltonian system with $d$ degree of freedom has
$d$ integrals in
involution, and the $d$ parameters in the general solution are given as the
levels of these
integrals.  With an invariant function of (1), we also have:
\enddemo

\proclaim{Proposition 2}  Let $\xi (x)$ be an invariant function, i.e., ${d\xi
\over dt}=0$, and $X_S
$ be a symmetry of (1).  Then $\varphi (\xi (x))X_S$ with an arbitrary scalar
function $\varphi$
of $\xi$ is also a symmetry of (1).
\endproclaim

The proof is straight-forward.  This proposition is useful for constructing
higher order symmetries
as we will see below.

For the case of linear system, $F(x)=Ax$ with $A\in M_N(\Bbb C)$, the symmetry
is just a solution of
the system.  If we assume the eigenvalues of $A$ to be all distinct, i.e.,
$\lambda_i\not= \lambda_j$
for $i\not= j$, then we have:

\proclaim{Proposition 3}  The vectors $A^nx$, for $n=0,\dots ,N-1$, give $N$
linearly independent
symmetries of the linear system $F(x)=Ax$ in (1) with the above condition for
$A$.
\endproclaim

\demo{Proof}  Let $e_\ell$ be the eigenvectors of $A$ with eigenvalue
$\lambda_\ell$, i.e., $Ae_\ell
=\lambda_\ell e_\ell$.  Since $A$ is diagonalizable, $e_\ell$'s are linearly
independent.  The
equations $\sum\limits^N_{i=1}a_iA^{i-1}e_\ell =0$, for $\ell =1,\dots ,N$,
give
$$\eqalignno{\left[\matrix 1 &\lambda _1 &\dots &\lambda^{N-1}_1\\
1 &\lambda_2 &\dots &\lambda^{N-1}_2\\
\vdots &\vdots &&\vdots\\
1 &\lambda_n &\dots &\lambda^{N-1}_N
\endmatrix\right]\left[\matrix a_1\\ a_2\\ \vdots\\ a_N\endmatrix\right] &=
0\,. &(18)\cr}$$
This equation has only trivial solution $a_i=0$ for all $i$ since the
determinant of the coefficient
matrix is the Vandermonde determinant $\prod\limits_{i>j}(\lambda_i
-\lambda_j)$ which is not zero
by the assumption.\rbx
\enddemo

The (linear) symmetries given by polynomial of degree one are called the {\it
basic} symmetries.  The
Proposition 3 then implies that any linear symmetry can be written as a linear
combination of the
basic symmetries. Namely, for a matrix with distinct eigenvalues, we have
$$ Span\{A^n : 0\le n \le {N-1}\} = \{ B \in M_N(\Bbb C) : [ A, B] = 0 \}.
\eqno(19)$$
  In order to find nonlinear symmetry, especially of polynomial type, we look
for an invariant function of the system and use the Proposition 2,as will be
shown below.

\heading{4.}
\endheading

Hereafter we consider the system (4) with the linear part having distinct
eigenvalues.  Note that the set
of matrix with distinct eigenvalues is generic in $M_N(\Bbb C)$.  In this case,
we first note that
the resonant condition (5) leads to an invariant function of the linear system:

\proclaim{Propoisition 4}  If there exists a resonance, i.e., $\exists
\lambda_s=\sum\limits^N_{i=1}
n_i\lambda_i$ for some $n_i\in\Bbb Z_{\ge 0}$ with $|n|\ge 2$, then
a function
$$\eqalignno{\xi (x) &= x^{-1}_s\prod^N_{i=1} x^{n_i}_i\,, &(20)\cr}$$
with the coordinates $x_i$ on which $A=\diag (\lambda_1,\dots ,\lambda_N)$, is
an invariant function.
\endproclaim

The proof to show $d\xi (x)/dt = 0$ is straight-forward.  Here we remark that
if $n_s\not= 0$,
then (20) gives a monomial of degree $|n| -1$. Among all the invariant
functions of the form (20) for each independent $n\in\Bbb Z^N_{\ge 0}$, the
invariant with the minimum $|n|$ is called {\it primitive}.  From the
Proposition
2, we have:

\proclaim{Proposition 5}  The vector fields defined by
$$\eqalignno{X^m_{\ell} =S^m_{\ell}(x)\cdot{d\over dx} &:= \xi^m(x)(A^\ell
x)\cdot{d \over dx}\,,\quad \text{for}~~\left\{\matrix
m\in\Bbb Z^d_{\ge 0}\,,\\
\ell =0,1,\dots ,N-1,\endmatrix\right. &(21)\cr}$$
are symmetries of the linear system.  Here $d$ is the total number of primitive
invariant functions, $\{
\xi_i(x)\}^d_{i=1}$, and in the coordinates where $A=\diag (\lambda_1,\dots
,\lambda_N)$, the
invariant function $\xi (x)$ is given by
$$\eqalignno{&\left.\matrix
\xi_j(x)=x^{-1}_{s_j}\prod\limits^N_{i=1}x^{n^{(j)}_i}_i:=\prod\limits^N_{i
=1}x^{\overline{n}^{(j)}_i}_i\\
\xi (x) =
%% FOLLOWING LINE CANNOT BE BROKEN BEFORE 80 CHAR
\prod\limits^d_{j=1}(\xi_j(x))^{m_j}:=\prod\limits^N_{i=1}x^{M_i}_i\endmatrix\right\}
&(22)\cr}$$
where $\lambda_{s_j}=\sum\limits^N_{i=1}n^{(j)}_i\lambda_i$ with
$\sum\limits^N_{i=1}n^{(j)}_i\ge 2$
(resonant condition), and $M_i=\sum\limits^d_{j=1}\overline{n}^{(j)}_jm_j$ with
$\overline{n}^{(j)}_i
=n^{(j)}_i-\delta_{ij}$.
\endproclaim

Note that in the symmetries (21) the number $d$ is just the total number of
resonances with independent $n\in \Bbb Z^N_{\ge 0}$.  In the
formula (21), we remark that if $n^{(j)}_{s_j}\not= 0$, $X^m_{\ell} (x)$ gives
a polynomial symmetry
for any $m\in\Bbb Z^d_{\ge 0}$ and $\ell$.  However, if $n^{(j)}_{s_j}=0$, $X^
m_{\ell}(x)$ is not a
polynomial, but a certain linear combination of $X^m_{\ell}(x)$ with $m_j=1$
becomes a polynomial.
In the latter case, there is no higher degree polynomial symmetries with
respect to the index $m_j$ (i.e.,
$\xi_j(x)x_{s_j}$ is polynomial, but $\xi_j^{m_j}x_{s_j}$ for $m_j>1$ is not).
We now show that
any polynomial symmetry can be given by a linear combination of (21):

\proclaim{Lemma 6}  If $X_S = S(x)\cdot {d\over dx}$ is a polynomial symmetry,
then there exists an invariant function $\xi
(x)$ of (20) and
$$\eqalignno{S(x) &= \xi (x)\sum^{N-1}_{i=0}C_{i}A^ix &(23)\cr}$$
for some constants $C_i$, for $i=0,\dots ,N-1$.
\endproclaim

\demo{Proof}  We first note that $S(x)$ is a solution of the linear equation,
i.e.,
$$\eqalignno{{d \over dt} S(x) &= AS(x)\,. &(24)\cr}$$
Since $A$ is diagonalizable, we take $A=\diag (\lambda_1,\dots ,\lambda_N)$
without loss of
generality.  Then (24) leads to the equations for the component $s_j(x)$ of
$S(x)$,
$$\eqalign{{d \over dt}s_j(x) &={dx \over dt}\cdot{d \over dx}
s_j(x)=(Ax)\cdot{d \over dx}s_j(x)\cr
&= \sum^N_{i=1}\lambda_ix_i{\partial \over \partial x_i}s_j(x)\cr
&= \lambda_js_j(x)\,,~~~\qquad~~~\text{for}~~j=1,\dots ,N\,.\cr}$$
This implies that  if $s_j(x)$ is a polynomial, it is a monomial in the form,
$$\eqalignno{s_j(x) &= \prod^N_{i=1}x^{n^{(j)}_i}_i\,. &(25)\cr}$$
Substituting (25) into (24), we obtain
$$\left(\sum^N_{i=1}n^{(j)}_i\lambda_i-\lambda_j\right) s_j(x) = 0\,.$$
This is just the resonant condition (6).  Namely the polynomial symmetry
results from the resonant
condition.  Then, from the Proposition 4, there exists an invariant function of
the form (20),
from which we have
$$\eqalignno{s_j(x) &= \xi_j(x)x_j=\left( x^{-1}_j
\prod^N_{i=1}x^{n^{(j)}_i}_i\right) x_j\,.
&(26)\cr}$$
Since the set of the basic symmetries $\{ A^n x\}^{N-1}_{n=0}$ gives a basis of
$\Bbb C^N$, a
symmetry $S(x)=(0,\dots,0,s_j(x),0,\dots ,0)^t$ can be expressed as a linear
combination of the basic
symmetries.  This completes the proof.\rbx
\enddemo

It should be noted that the symmetry given by (25) with zeros for other
component, $S(x)=(0,\dots ,
0,s_j(x),0,\dots ,0)^t$, is nothing but a resonant vector of (8) for $A=\diag
(\lambda_1,\dots ,
\lambda_N)$.  Therefore, we have the main theorem of this paper on the normal
form of (4):

\proclaim{Theorem 7}  The normal form (7) can be expressed as
$$\eqalignno{{dy \over dt} &= Ay+\sum_{\vert m\vert\ge 1}~\sum_{0\le\ell\le
N-1}C^m_\ell S^m_\ell
(y)\,, &(27)\cr}$$
where $S^m_\ell (y)$ for $m\in\Bbb Z^d_{\ge 0}$ and $\ell =0,1,2,\dots ,N-1$
are given by (21),
$C^m_\ell$ are constants determined uniquely from the original system (4), and
$\vert m\vert =\sum\limits^d_{i=1}m_i$.
\endproclaim

Let us illustrate the theorem by taking a nonlinear system of $\Bbb R^2$ near
an elliptic point as a
simple example.  Namely we consider the system,
$$\eqalignno{{d \over dt}{x_1 \choose x_2} &= \pmatrix 0 &1\\ -1
&0\endpmatrix~{x_1 \choose x_2}
+F^{(2)}(x_1,x_2)+\cdots\,. &(28)\cr}$$
The primitive invariant polynomial of the linear part of this sytem is $\xi
(x_1,x_2)=x^2_1+x^2_2$, and the
symmetries (21) are given by
$$S^m_0 (x) = (x^2_1+x^2_2)^m{x_1 \choose x_2}\,,\quad  and \quad  S^m_1 (x) =
(x^2_1+x^2_2)^m{x_2 \choose -x_1}\,.\eqno (29)$$
  Introducing the complex variable $z=x_1+ix_2$, the normal
form (27) can be written in the following familiar form [1],
$$\eqalignno{{dz \over dt} &= iz+\sum^\infty_{m=1}C^m\vert z\vert^{2m}z
&(30)\cr}$$
where $C^m=C^m_0-iC^m_1$.

\heading{5.}
\endheading

We now discuss the algebra generated by the symmetries (21).  A set of vector
fields naturally
forms an infinite dimensional Lie algebra.  Here we show that the symmetry
algebra gives an infinite
dimensional graded Lie subalgebra of the vector fields.  To do this, we compute
the commutator of
$X^m_\ell$ and $X^{m'}_{\ell '}$ by writing these in the coordinates with
$A=\diag (\lambda_1,
\dots ,\lambda_N)$, i.e.,
$$\eqalignno{X^m_{\ell} &=\left(\prod^N_{i=1}x^{M_i}_i\right)
\sum^N_{i=1}\lambda^\ell_ix_i{\partial
\over \partial x_i}\,. &(31)\cr}$$
Then it is immediate that we have:

\proclaim{Theorem 8}  The set of the symmetries (21) of the linear system forms
an infinite dimensional
Lie algebra with the commutator,
$$\eqalignno{[X^m_{\ell}, X^{m'}_{\ell '}] &=
\left(\sum^N_{i=1}M_i'\lambda^\ell_i\right) X^{m+m'}_{\ell
'} - \left(\sum^N_{i=1}M_i\lambda^{\ell '}_i\right) X^{m+m'}_\ell\,,
&(32)\cr}$$
where $M'_i=\sum\limits^d_{j=1}\overline{n}^{(j)}_i m'_j$.
\endproclaim

We have several remarks on the relation (32):

\remark{Remarks}  a)  This algebra is graded by the integer $m\in\Bbb Z^d_{\ge
0}$ of the degree
of the invariant functions $\xi_j(x)$; $\Bbb Z^d$-gradation.

b)  Since $X^0_1=(Ax)\cdot{d \over dx}$ is the original vector field, $X^0_1$
gives a center
of this algebra (recall $\sum\limits^N_{i=1}M_i\lambda_i=0$).

c)  The set of $X^m_0$ forms a classical Virasoro algebra (Witt algebra), i.e.
$$\eqalignno{[X^m_0, X^{m'}_0] &= \left(\sum M'_i - \sum M_i\right)
X^{m+m'}_0\cr
&= \left(\sum^d_{j=1}N^{(j)}(m'_j-m_j)\right) X^{m+m'}_0\,, &(33)\cr}$$
where $N^{(j)}:=\sum\limits^N_{i=1}\overline{n}^{(j)}_i$.
\endremark

Thus nonlinear system having resonances naturally carries an infinite
dimensional Lie algebra of the
symmetries, while linear system having no-resonance has only a finite
dimensional Abelian algebra of
the basic symmetries.

For the example (28), we have an algebra of Kac-Moody ($\widetilde{u(1)}$) and
Virasoro (Witt) type as the symmetry algebra,
$$\eqalignno{ [X^m_0, X^{m'}_0] &= 2(m'-m) X^{m+m'}_0,  \cr
[X^m_0, X^{m'}_1] &= 2m' X^{m+m'}_1, \cr
[X^m_1, X^{m'}_1] &= 0\,.   &(34)\cr}$$

\heading{6.}
\endheading

As a consequence of the previous result, we here show that the method of
multiple scales with more than two scaling variables fails in general.

Let $t_n$ be the scaling variables defined as
$$ t_n = {\epsilon}^n t, \quad  for \quad  n=0, 1, 2, \dots , \eqno (35)$$
where $\epsilon > 0$ is a small scaling parameter for the variable $x$.
Then, changing $x$ to $\epsilon x$, (4) becomes,
$$ {\partial  x \over \partial t_0} - Ax = \sum_{n=1}^{\infty} {\epsilon}^n
\left(
F^{(n+1)}(x) - {\partial x \over \partial t_n} \right). \quad \eqno (36)$$
Now, expanding $x$ in the power series of $\epsilon$, equivalently in
the homogeneous polynomials, i.e.
$$ x = x^{(1)} + \epsilon x^{(2)} + {\epsilon}^2 x^{(3)} + \cdots , \eqno
(37)$$
where $x^{(\ell)}$ is the homogeneous polynomial of degree $\ell$.
Substituting (37) into (36), we have at order 1,
$$ {\partial x^{(1)} \over \partial t_0} - Ax^{(1)} = 0 \quad \eqno (38)$$
which is the linearlized equation of (1), and in the case of $A$ having
distinct eigenvalues, all the polynomial solutions of (38) are given
by the symmetries (21).

At order $\epsilon$, we have
$$ {\partial x^{(2)} \over \partial t_0} - Ax^{(2)} =
F^{(2)}(x^{(1)}) - {\partial x^{(1)} \over \partial t_1}.  \quad \eqno (39)$$
Now as a standard procedure of perturbation method, we eliminate  secular
terms in $F^{(2)}$ by choosing an appropriate equation for ${\partial x^{(1)}
\over \partial t_1}$. The secular term is given by the homogeneous solution
of (39), that is, the symmetry of (38).  Thus we choose ${\partial x^{(1)}
\over \partial t_1}$ to be
$$ {\partial x^{(1)} \over \partial t_1} = S^{(2)}(x^{(1)}) =
F^{(2)}(x^{(1)}) \cap  kerL_A.  \quad \eqno (40)$$
In a similar manner, we have ${\partial x^{(1)} \over \partial t_n} =
S^{(n+1)}(x^{(1)}) \in  kerL_A$, and the equation for $x^{(1)}$ in
the original time variable becomes
$$ {dx^{(1)} \over dt} = \sum_{n=0}^{\infty} {\epsilon}^n {\partial x^{(1)}
\over \partial t_n} = Ax^{(1)} + \sum_{n=1}^{\infty} {\epsilon}^n S^{(n+1)}
(x^{(1)}).  \quad  \eqno (41)$$
This is nothing but the normal form (27) with $y=x^{(1)}$.  It is however
important to note that the equations of higher orders, ${\partial x^{(1)}
\over \partial t_n} = S^{(n+1)}(x^{(1)})$, are not compatible in general
 , as was
shown before, i.e. $[X_{S^{(n+1)}}, X_{S^{(m+1)}}] \ne 0$, or equivalently
$$ {\partial^2 x^{(1)} \over \partial t_n \partial t_m} \ne {\partial^2
x^{(1)} \over \partial t_m \partial t_n}, \quad for \quad  n \ne m.  \eqno
(42)$$
This implies that the method of multiple scales, where the scaling
vriables $t_n$ are considered to be independent, fails in general.  Note
however that the normal form (27) makes sense.

As a final remark, we mention that a similar result can be obtained
for a Hamiltonian system where the symmetry algebra is given by a Poisson
algebra over resonant polynomials.

\vskip 1cm

\centerline{\bf Acknowledgement}
\bigskip

The author wishes to thank Toni Degasperis, Allan Fordy, Thomas Kappeler,
and Kuo-chi Wu for valuable discussion.  The work is partially supported
by NSF Grant DMS-9109041.

\vfil\eject

\centerline{\bf References}
\bigskip

\list{[1]}{V. I. Arnold, Geometrical Methods in the Theory of Ordinary
Differential Equations,
(Springer-Verlag, New York, 1983).}

\enddocument